\documentclass[aps,prb,twocolumn,showpacs,superscriptaddress]{revtex4}
\usepackage{epsfig}
\usepackage{graphicx}
\usepackage{amsfonts}
\usepackage[figuresright]{rotating}
\usepackage{amssymb}
\usepackage{amsmath}

\font\tenmib=cmmib10
\font\eightmib=cmmib10 scaled 800
\font\sixmib=cmmib10 scaled 667
\newfam\mibfam

\def\mib{\fam\mibfam\tenmib}
\textfont\mibfam=\tenmib
\scriptfont\mibfam=\eightmib
\scriptscriptfont\mibfam=\sixmib
\mathchardef\alpha="710B
\mathchardef\beta="710C
\mathchardef\gamma="710D
\mathchardef\delta="710E
\mathchardef\epsilon="710F
\mathchardef\zeta="7110
\mathchardef\eta="7111
\mathchardef\theta="7112
\mathchardef\kappa="7114
\mathchardef\lambda="7115
\mathchardef\mu="7116
\mathchardef\nu="7117
\mathchardef\xi="7118
\mathchardef\pi="7119
\mathchardef\rho="711A
\mathchardef\sigma="711B
\mathchardef\tau="711C
\mathchardef\phi="711E
\mathchardef\chi="711F
\mathchardef\psi="7120
\mathchardef\omega="7121
\mathchardef\varepsilon="7122
\mathchardef\vartheta="7123
\mathchardef\varrho="7125
\mathchardef\varphi="7127
\mathchardef\sGamma="7100
\mathchardef\sDelta="7101
\mathchardef\sTheta="7102
\mathchardef\sLambda="7103
\mathchardef\sXi="7104
\mathchardef\sPi="7105
\mathchardef\sSigma="7106
\mathchardef\sUpsilon="7107
\mathchardef\sPhi="7108
\mathchardef\sPsi="7109
\mathchardef\sOmega="710A

\def\ssr#1{{\sss{\rm #1}}}
\def\sss#1{{\scriptscriptstyle #1}}

\def\nd{^{\vphantom{\dagger}}}
\def\ns{^{\vphantom{*}}}
\def\yd{^\dagger}
\def\frac#1#2{{\textstyle{#1 \over #2}}}

\def\ie{{\it i.e.\/}}

\def\pz{{\partial}}
\def\half{\frac{1}{2}}
\def\vF{v\ns_\ssr{F}}
\def\bfk{{\mib k}}
\def\bfK{{\mib K}}

\def\bfx{{\mib x}}
\def\bfsigma{{\mib \sigma}}
\def\ve{\varepsilon}
\def\vF{v\ns_\ssr{F}}
\def\bnabla{\boldsymbol{\nabla}}
\def\xhat{{\hat{\mib x}}}
\def\yhat{{\hat{\mib y}}}
\def\nhat{{\hat{\mib n}}}
\def\cN{{\cal N}}

\def\bfJ{{\mib J}}

\begin{document}

\title{Dirac Spectrum in Piecewise Constant One-Dimensional Potentials}
\author{D. P. Arovas}
\affiliation{Dept. of Physics, University of California, San Diego, CA 92093}
\author{L. Brey}
\affiliation{Instituto de Ciencia de Materiales de Madrid
(CSIC), Cantoblanco 28049, Spain}
\author{H. A. Fertig}
\affiliation{Department of Physics, Indiana University, Bloomington, IN 47405}
\author{Eun-Ah Kim}
\affiliation{Department of Physics, Cornell University, Ithaca, NY 14853}
\author{K. Ziegler}
\affiliation{Institut f\"ur Physik, Universit\"at Augsburg
D-86135 Augsburg, Germany}

\begin{abstract}
We study the electronic states of graphene in piecewise constant potentials
using the continuum Dirac equation appropriate at low energies, and
a transfer matrix method.  For superlattice
potentials, we identify patterns of induced Dirac points which are present
throughout the band structure, and verify for the special case
of a particle-hole symmetric potential their presence at zero energy.
We also consider the cases of a single trench and a $p$-$n$ junction
embedded in neutral graphene, which are shown to support confined
states.  An analysis of conductance across these structures demonstrates
that these confined states create quantum interference effects which
evidence their presence.
\end{abstract}

\maketitle
\section{Introduction}
Graphene, a single layer of carbon atoms laid out in a honeycomb lattice, is one of
the most interesting electronic systems to become available the last few years\cite{Geim07_nmt,neto:109}.
It differs from conventional two dimensional electron gas (2DEG) systems in that the
low energy physics is governed by a Dirac Hamiltonian rather than the
more common form for semiconductors,
characterized by an effective mass and a band gap.  Because of its
unusual properties, the possibility of exploiting it for nanoscale applications
has become an area of intense investigation.

Nanophysics in graphene may be investigated in systems with very small
physical size scales, such as quantum dots \cite{dots_com} or quantum
rings \cite{molitor09,luo09,abergel08,bahamon08,potas09,wurm09}.  More commonly,
manipulation at the nanoscale is accomplished
by the application of electric fields via gate geometries, subjecting the system
to potentials varying on a very short length scale.
Much recent work  has focused
on $p$-$n$ junctions \cite{stander09,young09,cheianov07,zhang08,beenakker08}
and $p$-$n$-$p$ junctions \cite{velasco09,pereira06},
whose behavior in graphene is distinct from that of conventional 2DEG's
because of the absence of a gap in the spectrum.
Very recently, studies of graphene in superlattice potentials
have demonstrated the possibility of ``band structure engineering''
of graphene \cite{pletikosic09,tiwari09}.  In particular, for one-dimensional
superlattices \cite{esmailpour08,park08},
effective potentials may be designed by which the number of Dirac
points at the Fermi energy can be artificially manipulated \cite{pereira07,barbier08,barbier09,bliokh09,brey09,park09}.

In this work, we study graphene systems in the presence of one-dimensional potentials
that are piecewise constant.  Our approach models graphene in terms of the Dirac equation, which is
valid for potentials that vary slowly on the lattice scale, and for which intervalley scattering may be ignored.
The allowed energy states for such systems may be found by a transfer matrix method described below.
For a one-dimensional superlattice, such problems are reminiscent of dynamical systems
with periodically varying parameters.  We find that the resulting
band structure generically supports a collection of Dirac points at a variety of energies.
Dirac points generally appear at energies satisfying $\varepsilon \equiv Ed/\hbar\vF = n\pi$ ($E=$
electron energy, $d=$ superlattice period, $\vF=$ Fermi velocity) for $n$ an integer,
at $q=0$, where $q$ is the momentum perpendicular to the superlattice axis.
For particle-hole symmetric potentials, several Dirac points may appear at $\varepsilon=0$
for non-vanishing values of $q$ \cite{pereira07,barbier08,barbier09,bliokh09,brey09,park09,barbier10}.
Generically, other Dirac points are present for $\varepsilon,~q\ne 0$.

We also study geometries with a single $p$ or $n$ region, or with a $p$-$n$ junction
region, embedded in a neutral graphene background.  We refer to these as
a single trench geometry and an embedded $p$-$n$ junction, respectively.
The trench geometry supports localized solutions,
at energies for which the electron states are evanescent in the
neutral graphene for a given $k_y \ne 0$.  These states disperse
as a function of $k_y$ such that a finite number of them crosses zero energy.
The embedded $p$-$n$ junction supports localized solutions as well,
which may be interpreted in terms of a pair of overlaid trench spectra.
If the two trenches are particle-hole conjugates, the individual trench
spectra cross zero energy at the same value of $k_y$; however, because of
tunneling between the trenches these become avoided crossings rather
than Dirac points.  At large $k_y$ the gaps at the avoided crossings
may be quite small.

Finally we consider transport across these systems within the Landauer
formalism, for which the transfer matrix method is easily adapted to
yield the conductance of the system.  We demonstrate that the
conductance has distinctive signatures indicative of the confined states
which may be contained at non-vanishing $k_y$ for a given embedded structure.

This article is organized as follows. In Section \ref{periodic} we express
the problem of a Dirac particle in a periodic potential in terms of
a path-ordered product, and find the band structure of the simplest case
of alternating positive and negative potential regions.  Section \ref{transfer_matrix}
develops the transfer matrix method for handling more general piecewise
constant potentials.  In Section \ref{localized} we apply the method to
compute the spectrum of localized solutions in trench and embedded $p$-$n$
junction geometries.  Section \ref{conductance} applies the results of
the transfer matrix calculations to compute the conductance of these system.
Finally, we conclude with a summary in Section \ref{summary}.

\section{Periodic potential}
\label{periodic}
In the vicinity of a Dirac point at wavevector $\bfK$, we write the wavevector $\bfk=\bfK+\Delta\bfk$, and
the two component wavefunction as $\psi\equiv e^{i\bfK\cdot\bfx} \, \big ( {u \atop v} \big )$.  Assuming
a scalar potential $V(x)$ which is independent of the coordinate $y$, the wavefunction $\phi=\big ( {u \atop v} \big )$
may be chosen an eigenstate of the Hamiltonian ${\widetilde H}=e^{-i\bfK\cdot\bfx}\, H\,e^{i\bfK\cdot\bfx}\,$:
\begin{equation}
{\widetilde H}=\begin{pmatrix} V(x) & \hbar\vF (-i\pz\nd_x + iq) \\
\hbar\vF (-i\pz\nd_x - iq) & V(x) \end{pmatrix}\ ,
\end{equation}
where $q=\Delta k\ns_y$.
The eigenvalue equation for $\phi$ may then be recast in matrix form as $\pz\nd_x\phi=M(x)\,\phi$, where
\begin{equation}
M(x)=\begin{pmatrix} -q & i\kappa(x) \\ i\kappa(x) & q \end{pmatrix}
\label{meqn}
\end{equation}
and
\begin{equation}
\kappa(x)= {E-V(x)\over\hbar\vF}\ .
\end{equation}
The solution is
\begin{equation}
\begin{pmatrix} u(x) \\ v(x) \end{pmatrix}={\cal P}\exp\Bigg(\int\limits_0^x\!\!dx'\>M(x')\Bigg)\begin{pmatrix} u(0) \\ v(0) \end{pmatrix}\ ,
\end{equation}
where {\cal P} denotes the path ordering operator, which places smaller values of $x$ to the right.

We presume $V(x)=V(x+d)$ is periodic, in which case the Bloch condition becomes
\begin{equation}
\begin{pmatrix} u(d) \\ v(d) \end{pmatrix}=e^{i\theta\ns_x} \begin{pmatrix} u(0) \\ v(0) \end{pmatrix}\ ,
\end{equation}
where $\theta\ns_x$ is the Bloch phase.  This leads to the condition
\begin{equation}
\textsf{det}\big[ \Lambda - e^{i\theta\ns_x}\big]=0\ ,
\end{equation}
where
\begin{equation}
\Lambda={\cal P}\exp\Bigg(\int\limits_0^d\!\!dx\>M(x)\Bigg)\ .
\end{equation}
Note that $M\yd=-\sigma^x M \sigma^x$, where $\sigma^x$ is the Pauli matrix, and that $\textsf{Tr}\,M=0$.
Since $\textsf{det}(\Lambda)=1$, the eigenvalues
of $\Lambda$ are given by $\zeta=\half T \pm\sqrt{T^2-4}$, where $T=\textsf{Tr}\,(\Lambda)$.  The energy bands, where the Bloch
condition can be satisfied, correspond to the regime $|T|\le 2$.  For $|T|>2$ the wavefunctions grow exponentially and are not
normalizable.  The mathematical structure here is common to a variety of problems involving linear equations with periodic coefficients\cite{ArnoldODE},
such as the classical mechanics setting of a pendulum with periodically varying length.

For a continuous periodic function $V(x)$ the path ordered exponential is not analytically tractable.  However, many of the essential features
are reproduced by considering a piecewise constant $V(x)$ of the form
\begin{equation}
V(x)=\begin{cases} V\ns_1 & {\rm if} \quad 0 \le x < d\ns_1 \\
V\ns_2 & {\rm if} \quad d\ns_1 \le x < d\ns_1+d\ns_2\ .\end{cases}
\end{equation}
We then have $\Lambda=\Lambda\ns_2\,\Lambda\ns_1$, where $\Lambda\ns_j=\exp(M\ns_j d\ns_j )$, where
the matrices $M\ns_j$ ($j=1,2$) are as in Eq. \ref{meqn}, with $\kappa\ns_j\equiv (E-V\ns_j)/\hbar\vF$ in
the off-diagonal elements.  The period of $V(x)$ is $d\ns_1+d\ns_2$.  We then have
\begin{equation}
\Lambda\ns_j=\cos\alpha\ns_j \cdot {\mathbb I} + {\sin \alpha\ns_j\over\alpha_j}\cdot  M\ns_j d\nd_j\ ,
\end{equation}
with $\alpha\ns_j=d\ns_j\sqrt{\kappa_j^2-q^2\>}$.  The resulting trace is
\begin{equation}
T=2\cos\alpha\ns_1\cos\alpha\ns_2 + 2\,d\ns_1d\ns_2\,\big(q^2-\kappa\ns_1\kappa\ns_2\big)
\cdot{\sin\alpha\ns_1\sin\alpha\ns_2\over\alpha\ns_1\,\alpha\ns_2}\ .
\label{treqn}
\end{equation}
The Bloch condition is satisfied when $|T|\le 2$.  It is convenient to define the dimensionless quantities
\begin{align}
\ve & = \big(E-\langle V\rangle\big)\cdot{d\ns_1+d\ns_2\over\hbar\vF}\\
\omega & =\half (V\ns_1-V\ns_2)\cdot {d\ns_1+d\ns_2\over\hbar\vF} \ ,
\end{align}
where $\langle V\rangle=(d\ns_1 V\ns_1 + d\ns_2 V\ns_2)/(d\ns_1+d\ns_2)$ is the average potential.  We also define
the dimensionless wavevector $\theta\ns_y=q \,(d\ns_1+d\ns_2)$.
When $V$ is constant, one finds $T=2\cos\sqrt{\ve^2-\theta_y^2\,}$, and the Bloch condition is satisfied provided $|E-V|>\hbar\vF |q|$.
This, of course, is the condition that the energy $E$ lies within the Dirac cone.

Next, consider $V\ns_1\ne V\ns_2$, but with $q=0$.  The trace is then $T=2\cos\ve$, and there are allowed states at all
energies.  The Bloch condition is satisfied marginally when $\ve=n\pi$,
corresponding to the touching of two energy bands which disperse with $q$, \ie\ to the vanishing
of a band gap as $q\to 0$.   Indeed this result holds for arbitrary $V(x)$, since $M(x)=\kappa(x)\,\sigma^x$ in this case,
hence $\big[M(x),M(x')\big]=0$ and the path ordering is superfluous.

\begin{figure}[t]
\begin{center}
\includegraphics[width=0.45\textwidth]{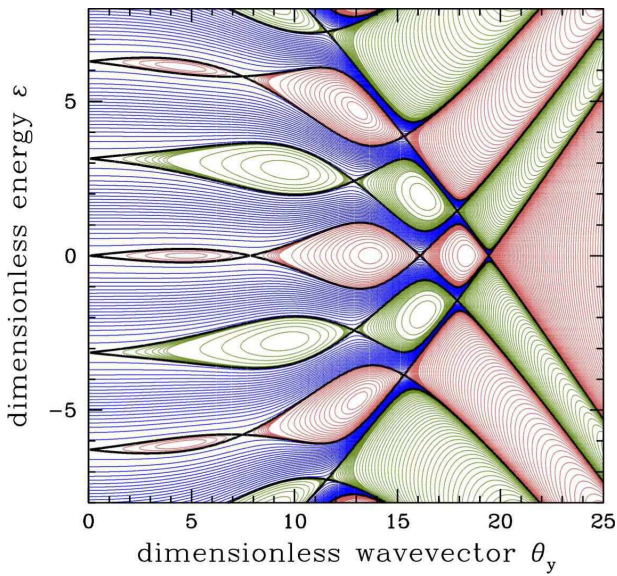}
\end{center}
\caption{(color online) Contour plot of $T=\textsf{Tr}\,(\Lambda)$ for $d_1=d_2$ and $\omega=\frac{13}{2}\pi$.
Red regions correspond to $T>2$, and olive regions to $T<-2$.  The energy bands are identified with the blue
regions, where $|T|<2$.}
\label{bands_01}
\end{figure}

Now consider the case $d\ns_1=d\ns_2$.  The spectrum then exhibits a particle-hole symmetry relative to $E=\langle V\rangle$.
When $E=\langle V \rangle $, \ie\ $\ve=0$, we have
\begin{equation}
T=2 + \left({2\theta\ns_y\,\sin\half \sqrt{\omega^2-\theta_y^2 \> } \over\sqrt{\omega^2-\theta_y^2\>}}\right)^{\!\!2}\ .
\end{equation}
Clearly $T\ge 2$, with the marginal condition $T=2$ pertaining when
\begin{equation}
\theta\ns_y=\pm\sqrt{\omega^2-4\pi^2 n^2\,}\ ,
\end{equation}
where $n$ is an integer, as well as a solution at $\theta\ns_y=0$.  Since $q$ is real, there is a maximum allowed value for $n$:
\begin{equation}
n\ns_{\rm max}=\bigg[{ \omega \over 2\pi}\bigg]\ ,
\end{equation}
where the brackets indicate the greatest integer function.  Thus, two new Dirac points open up each time $\omega$
increases by $2\pi$, and there are $2n\ns_{\rm max}\!+\!1$ Dirac points, including the one at $\theta\ns_y=0$.
Similar results have previously been obtained by other methods \cite{park09,brey09}.

\begin{figure}[t]
\begin{center}
\includegraphics[width=0.45\textwidth]{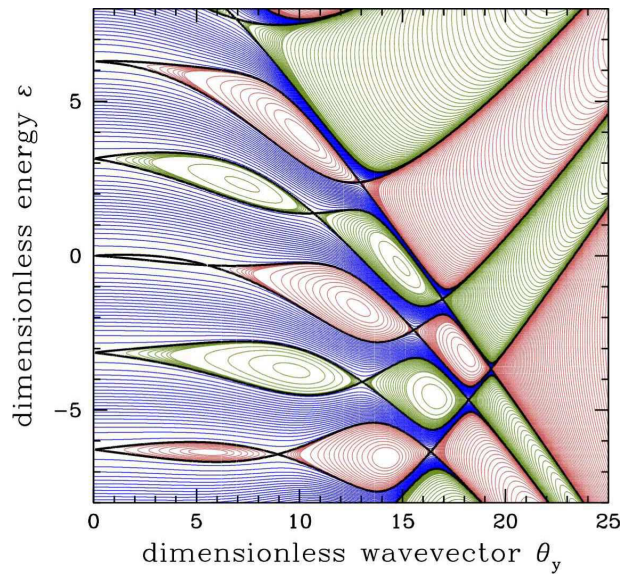}
\end{center}
\caption{(color online) Contour plot of $T=\textsf{Tr}\,(\Lambda)$ for $\omega=\frac{13}{2}\pi$ and $d_1=\frac{2}{3}\,d_2$.}
\label{bands_02}
\end{figure}

\begin{figure}[t]
\begin{center}
\includegraphics[width=0.45\textwidth]{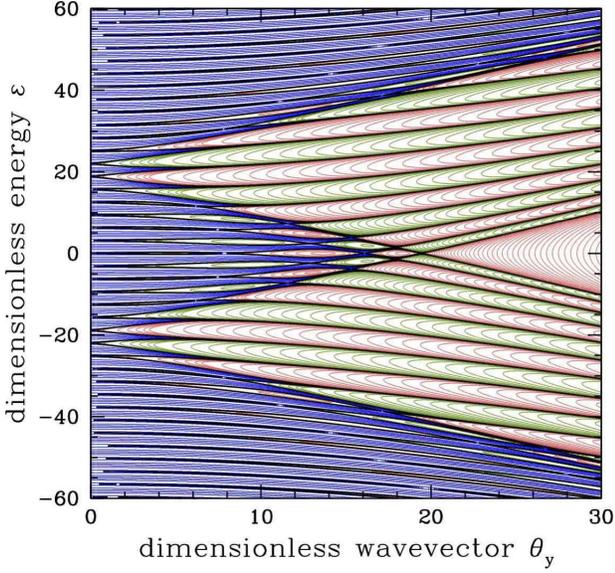}
\end{center}
\caption{(color online) Contour plot of $T=\textsf{Tr}\,(\Lambda)$ for $\omega_0=\frac{13}{2}\pi$ and $d_1=d_2$.}
\label{bands_03}
\end{figure}

In Figs. \ref{bands_01} and \ref{bands_02} we plot the contours of the function $T(\ve,\theta\ns_y)$ for $\omega=\frac{13}{2}\pi$
for $d\ns_1=d\ns_2$ and $d\ns_1=\frac{2}{3}\,d\ns_2$, respectively.  Note the particle hole symmetry present in the former case, which is broken
when $d\ns_1\ne d\ns_2$.  Note also the band touchings at $\ve=n\pi$ for $\theta\ns_y=0$ in both cases.  In Fig. \ref{bands_03} we again plot the
band structure as in Fig. \ref{bands_01}, but showing the high energy structure as well.  At energy scales $|\ve|\gg 1$,
the spectrum may be understood as the intersection of the two Dirac cones $ | \ve + \omega | > \theta\ns_y$ and $ | \ve - \omega | > \theta\ns_y$.

\begin{figure}[t]
\begin{center}
\includegraphics[width=0.45\textwidth]{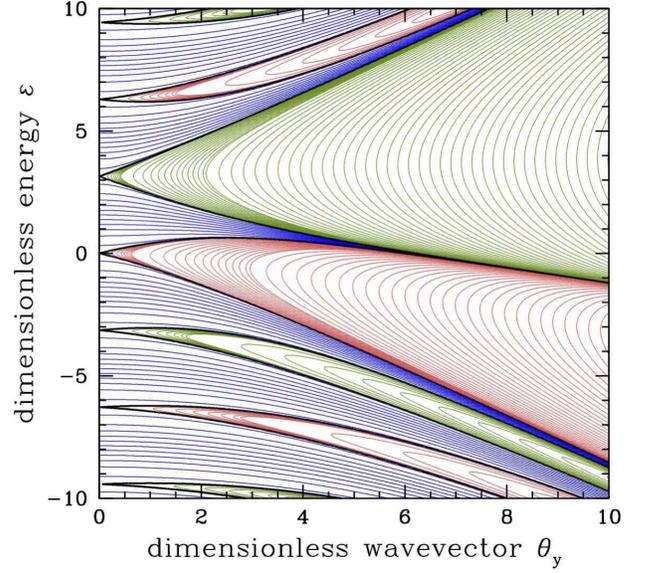}
\end{center}
\caption{(color online) Contour plot of $T=\textsf{Tr}\,(\Lambda)$ in the $d_2\to 0$ limit with
$\Omega_0 \equiv -V_2 d_2/\hbar v_\ssr{F}=\frac{3}{5}\pi$ held constant.}
\label{bands_06}
\end{figure}

As a limiting case of the model, consider $d\ns_2\to 0$ with $V\ns_2 d\ns_2=-\hbar\vF\Omega\ns_0$ finite.  The potential is then
\begin{equation}
V(x)=\Omega\ns_0\cdot{\hbar\vF\over d}\>\Bigg\{1-\sum_{n=-\infty}^\infty\delta\Big({x\over d}-n\Big)\Bigg\}\ ,
\end{equation}
\ie\ a Dirac comb with a uniform compensating background.  The trace in this case is given by
\begin{equation}
T=2\cos\Omega\ns_0\cos\alpha - 2\sin\Omega\ns_0\sin\alpha\cdot\bigg({\ve-\Omega\ns_0\over\alpha}\bigg)\ ,
\end{equation}
where $\alpha=\sqrt{(\ve-\Omega\ns_0)^2-\theta_y^2\,}$.   This limit is distinguished by the fact that all the Dirac points
of $T(\ve,\theta\ns_y)$ lie along the line $\theta\ns_y=0$, as shown in Fig. \ref{bands_06}.

\section{Transfer Matrices and Landauer Conductance Through Potential Slabs}
\label{transfer_matrix}
\begin{figure}[!t]
\centering\epsfig{file=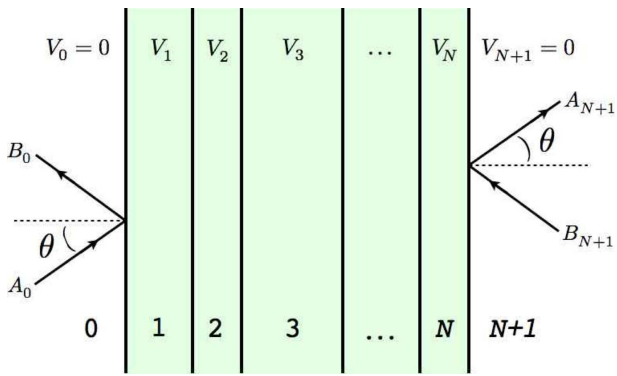,width=0.45\textwidth}
\caption{Transport through a series of potential slabs.  The potential $V(x)$ is piecewise constant, given by $V_j$ in the $j^\ssr{th}$ slab.}
\label{slabs}
\end{figure}

Consider a piecewise constant potential $V(x)$ as depicted in Fig. \ref{slabs} in which $V=V\ns_j$ over a slab of width $d\ns_j$, with
$j=1,\ldots, N$.  We presume the slabs are bounded by pristine regions of graphene with $V=0$, and we label these outer regions $0$ and $N+1$.
Wavevectors $k_j$ for each of the slabs may be defined by the relations
\begin{equation}
{E\over\hbar\vF}=\sigma\ns_0 \, k \ns_0= {V\ns_j\over\hbar\vF}+ \sigma\ns_j \, k\ns_j\ ,
\end{equation}
where each $k\ns_j\ge 0$ is nonnegative.  Here, $\sigma\ns_j=\pm 1$ is a binary variable in
each slab which indicates whether the energy $E$ is above (n-type) or below (p-type) the Dirac point,
which occurs at $E=V\ns_j$.  We take $\sigma\ns_0=\sigma\ns_{N+1}$ and $V\ns_0=V\ns_{N+1}=0$.

In each region $j$, the solution to the Dirac equation,
\begin{equation}
-i\hbar\vF\,\bfsigma\cdot\bnabla \psi + V\ns_j\psi = E\psi\ ,
\end{equation}
is either propagating or evanescent.  Due to translation invariance in the $y$-direction, we are free to fix the
wavevector $q$ along this axis; all solutions include a factor $e^{iqy}$.  The local Fermi wavevector $k\ns_j$ in
each region is related to the local density by $n\ns_j=k_j^2/4\pi$, and the sign of the carrier charge is $-\sigma\ns_j$.

If $k\ns_j > q$, the wavefunction in slab $j$ propagates as a plane wave along the $x$-direction as well.
The solution is of the form
\begin{align}
\psi\ns_j(x)&=\Bigg\{ {A\ns_j\over\sqrt{2}} \begin{pmatrix} 1 \\ \sigma\ns_j\,e^{i\theta\ns_j} \end{pmatrix} e^{ik\ns_j x \cos\theta\ns_j} \ + \label{WFprop}\\
&\qquad\qquad  {B\ns_j\over\sqrt{2}} \begin{pmatrix} 1 \\ -\sigma\ns_j\,e^{-i\theta\ns_j} \end{pmatrix} e^{-ik\ns_j x \cos\theta\ns_j} \Bigg\}
\ e^{ik\ns_j y\sin\theta\ns_j}\ , \nonumber
\end{align}
where $\theta\in \big[-\frac{\pi}{2},\frac{\pi}{2}\big]$.  Matching the $y$-dependence gives us the condition
\begin{equation}
q=k\ns_j\sin\theta\ns_j \quad \hbox{\rm if}\quad k\ns_j \ge q \ .
\end{equation}

To solve for the wavefunction everywhere we must match the value of each component of $\psi(x)$ at the boundaries between slabs.
Accordingly, from Eq. \ref{WFprop}, we define the matrix
\begin{equation}
M\ns_j=\begin{pmatrix} \frac{1}{\sqrt{2}} & \frac{1}{\sqrt{2}} \\ & \\ \frac{1}{\sqrt{2}} \, \sigma\ns_j\,e^{i\theta\ns_j} &
- \frac{1}{\sqrt{2}} \, \sigma\ns_j\,e^{-i\theta\ns_j} \end{pmatrix}\ ,
\end{equation}
which we shall invoke presently.  Free propagation across a slab is described by the transfer matrix
\begin{equation}
K\ns_j=\begin{pmatrix} e^{i k\ns_j d\ns_j\cos\theta\ns_j} & 0 \\ & \\ 0 & e^{-i k\ns_j d\ns_j\cos\theta\ns_j} \end{pmatrix}\ .
\end{equation}

If $k\ns_j < q$, the solution in slab $j$ has an evanescent (pure exponential) behavior along the $x$-axis:
\begin{align}
\psi\ns_j(x)&=\Bigg\{ A\ns_j \begin{pmatrix} \cos\big(\frac{\varphi\ns_j}{2}\big) \\ i\sigma\ns_j \sin\big(\frac{\varphi\ns_j}{2}\big) \end{pmatrix} e^{k\ns_j x \cot\varphi\ns_j } \  + \\
&\qquad\qquad  B\ns_j \begin{pmatrix} \sin\big(\frac{\varphi\ns_j}{2}\big) \\ i\sigma\ns_j \cos\big(\frac{\varphi\ns_j}{2}\big) \end{pmatrix} e^{-k\ns_j x \cot\varphi\ns_j} \Bigg\}
\ e^{i k\ns_j y\csc\varphi\ns_j} \ ,\nonumber
\end{align}
where $\varphi\ns_j\in \big[-\frac{\pi}{2},\frac{\pi}{2}\big]$.  Momentum conservation along $y$ requires
\begin{equation}
q=k\ns_j \csc\varphi\ns_j \quad  \hbox{\rm if}\quad k\ns_j \le q\ .
\end{equation}
For the evanescent slabs, we define
\begin{equation}
M\ns_j=\begin{pmatrix} \cos\big(\frac{\varphi\ns_j}{2}\big) & \sin\big(\frac{\varphi\ns_j}{2}\big) \\ & \\
i\sigma\ns_j\sin\big(\frac{\varphi\ns_j}{2}\big) & i\sigma\ns_j\cos\big(\frac{\varphi\ns_j}{2}\big) \end{pmatrix}\ ,
\end{equation}
and free propagation across an empty slab is described by
\begin{equation}
K\ns_j=\begin{pmatrix} e^{k\ns_j d\ns_j \cot\varphi\ns_j } & 0 \\ & \\ 0 & e^{-k\ns_j d\ns_j  \cot\varphi\ns_j } \end{pmatrix}\ .
\end{equation}

The equations guaranteeing continuity of the wavefunction at the slab boundaries are
\begin{equation}
M\ns_{j-1} \begin{pmatrix} A_{j-1}^\ssr{R} \\ \\ B_{j-1}^\ssr{R} \end{pmatrix} = M\ns_j \begin{pmatrix} A_j^\ssr{L} \\ \\ B_j^\ssr{L} \end{pmatrix} \ .
\end{equation}
Here, the superscripts R and L refer to the right and left boundaries of the slab.
They are related, within a given slab, by the free propagation transfer matrix,
\begin{equation}
\begin{pmatrix} A_j^\ssr{R} \\ \\ B_j^\ssr{R} \end{pmatrix} = K\ns_j \begin{pmatrix} A_j^\ssr{L} \\ \\ B_j^\ssr{L} \end{pmatrix}
\end{equation}

We can now compute the transfer matrix ${\cal M}$ for a general configuration of slabs, which is defined by the relation
\begin{equation}
\begin{pmatrix} A_{N+1}^\ssr{L} \\ \\ B_{N+1}^\ssr{L} \end{pmatrix} = {\cal M} \begin{pmatrix} A_0^\ssr{R} \\ \\ B_0^\ssr{R} \end{pmatrix} \ .
\end{equation}
In terms of the previously defined quantities, the total transfer matrix is
\begin{equation}
{\cal M}=M_{N+1}^{-1} \big(M\ns_N\,K\ns_N\, M_N^{-1}\big) \cdots \big(M\ns_1\,K\ns_1\, M_1^{-1}\big)  M\ns_0\ .
\label{transfermat}
\end{equation}
The recurring matrices $Q_j=M_jK_jM^{-1}_j$ may be parameterized by $\theta_j$ and $\chi_j=kd\cos \theta_j$ for
a propagating region, and by $\varphi_j$ and $\omega_j=kd\cot \varphi_j$ for an evanescent region,
as detailed in the Appendix.
As is well-known, the transfer matrix relates data on the left to data on the right of a given region.  Incoming and outgoing flux
amplitudes are related by the $S$-matrix,
\begin{equation}
\begin{pmatrix} A\ns_{N+1} \\ \\ B\ns_0 \end{pmatrix}
=\stackrel{{\cal S}}{\overbrace{ \begin{pmatrix} t && r' \\ && \\ r && t' \end{pmatrix} }}
\begin{pmatrix} A\ns_0 \\ \\ B\ns_{N+1}\end{pmatrix}.
\end{equation}
Appealing to the unitarity of ${\cal S}$, we then have
\begin{align}
{\cal M}\nd_{11} & = t^{*-1} & {\cal M}\nd_{12} &= r'\,t^{\prime-1}\\
{\cal M}\nd_{21} &= -t^{\prime-1}\,r & {\cal M}\nd_{22} &=  t^{\prime-1}\ .
\end{align}
As discussed in more detail below,
the dimensionless 2-terminal Landauer conductance, in units of $e^2/h$ per spin channel, is given by
\begin{equation}
T = |t|^2 = | {\cal M}\ns_{11} |^{-2}\ .
\end{equation}

\section{Localized Solutions of Embedded Geometries}
\label{localized}
We first consider situations where
the outer regions contain evanescent waves, so the wavefunction is localized within the slabs.
In this case, we must have $B\ns_0=0$ and $A\ns_{N+1}=0$, if $\cot\varphi>0$, which requires ${\cal M}\ns_{11}=0$;
for the case $\cot\varphi < 0$, we must have $A\ns_0=0$ and $B\ns_{N+1}=0$, leading to ${\cal M}\ns_{22}=0$.  This condition defines
the one-dimensional dispersion $E(q)$, where $q=k\ns_0\csc\varphi\ns_0=k\ns_j\sin\theta\ns_j$.

\begin{figure}[t]
\begin{center}
\includegraphics[width=0.45\textwidth]{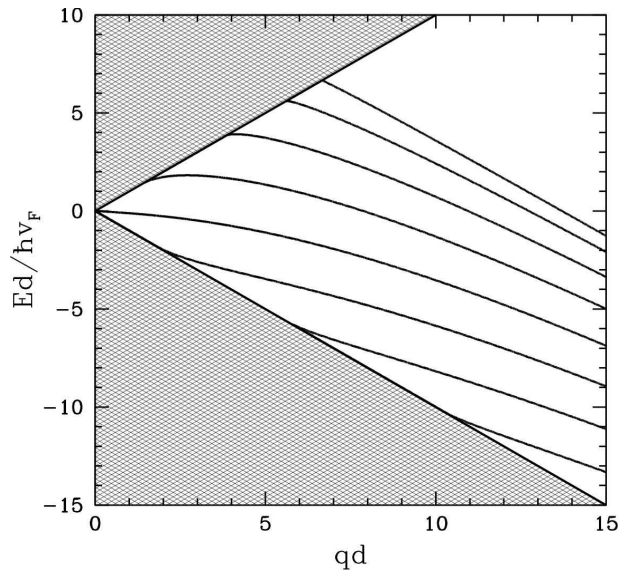}
\end{center}
\caption{Energy bands for a single trench of width $d$ and potential $V=14\,\hbar v_\ssr{F}/d$.  The hatched region depicts the Dirac cones in the bulk.
Solid lines show dispersions for bands of localized states in the gap between the upper and lower Dirac cones.}
\label{bands_07}
\end{figure}

\subsection{Single trench}
For the case where there is a single trench of width $d$, the condition ${\cal M}\ns_{11}=0$ is
\begin{equation}
\cos\varphi \ns_0\cos\theta\ns_1\cos\chi\ns_1+
\big(\sin\theta\ns_1 - \sigma\ns_0\,\sigma\ns_1\sin\varphi\ns_0\big) \sin\chi\ns_1 = 0
\end{equation}
or
\begin{equation}
\cos\varphi\ns_0\cos\varphi\ns_1\cosh\omega\ns_1
+\big(1 - \sigma\ns_0\,\sigma\ns_1\sin\varphi\ns_0\sin\varphi\ns_1\big) \sinh\omega\ns_1=0\ .
\end{equation}
The condition ${\cal M}\ns_{22}=0$ is obtained by reversing the sign of $\chi\ns_1$ or $\omega\ns_1$.
We now define the dimensionless potential $\nu\equiv Vd/\hbar\vF$, energy $\ve\equiv\sigma\ns_0 k\ns_0 d=\nu  + \sigma\ns_1 k\ns_1 d$,
and wavevector $\theta\ns_y=q d$.  The eigenvalue condition is then $F(\ve,\nu,\theta\ns_y)=0$, where
\begin{equation}
F(\ve,\nu,\theta\ns_y)=\Big(\theta_y^2-\ve\, (\ve-\nu)\Big)\cdot{\sin\chi\ns_1\over\chi\ns_1} + \sqrt{\theta_y^2-\ve^2\>}\>\cos\chi\ns_1
\end{equation}
and
\begin{equation}
\chi\ns_1=\sqrt{(\ve-\nu)^2-\theta_y^2\>}\ .
\end{equation}
This form for $F$ holds for either sign of $q$ and is extended to the regime $(\ve-\nu)^2 < \theta_y^2$ via analytic
continuation.  If we search for eigenvalues at $\ve=0$, we obtain the condition
\begin{equation}
{\chi\ns_1\over (\nu^2-\chi_1^2)^{1/2}}=-\tan\chi\ns_1\ ,
\end{equation}
a graphical analysis of which shows a new root emerging each time $u d$ passes through $n\pi$.
Thus, there are $\big[\nu/\pi\big]$ solutions with $\ve=0$ which lie at finite values of $\theta\ns_y$.   In Fig. \ref{bands_07},
for example, we have $\nu=14$ and thus there are $[14/\pi]=4$ bands which cross $\ve=0$ at finite $\theta\ns_y$.

\begin{figure}[t]
\begin{center}
\includegraphics[width=0.45\textwidth]{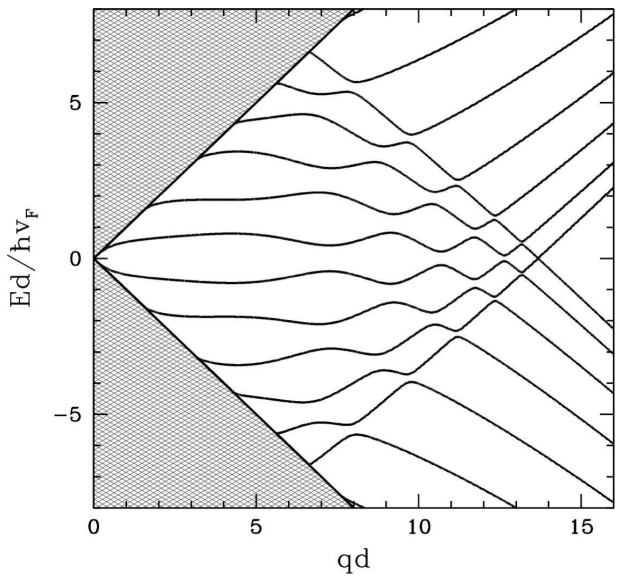}
\end{center}
\caption{Energy bands for a $p$-$n$ junction embedded in neutral graphene.  Both p and n regions have width $d$,  with
$V_1=-V_2= 14\,\hbar v_\ssr{F}/d$.   The hatched region depicts the Dirac cones in the bulk.
Solid lines show dispersions for bands of localized states in the gap between the upper and lower Dirac cones.}
\label{bands_08}
\end{figure}

\begin{figure}[t]
\begin{center}
\includegraphics[width=0.45\textwidth]{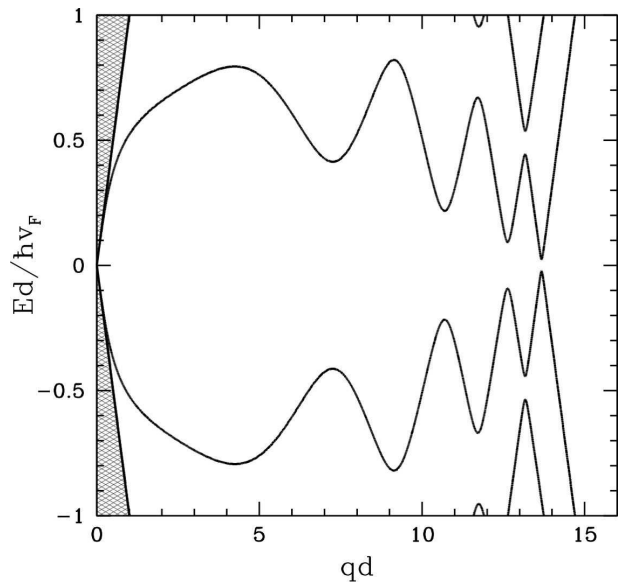}
\end{center}
\caption{Low energy detail for the embedded $p$-$n$ junction from Fig. \ref{bands_08}, showing that $E=0$ always lies within a gap between
subbands.}
\label{bands_09}
\end{figure}

\subsection{Embedded junction}
Consider next the case of an embedded junction with a double step,
\begin{equation}
V(x)=\begin{cases}
0 & \hbox{\rm if} \quad x <  0 \\
V\ns_1 & \hbox{\rm if} \quad 0 \le x < d\ns_1 \\
V\ns_2  & \hbox{\rm if} \quad d\ns_1 \le x < d\ns_1 + d\ns_2 \\
0 & \hbox{\rm if} \quad d\ns_1 + d\ns_2 \le x \ .
\end{cases}
\end{equation}
When $V\ns_1$ and $V\ns_2$ are of opposite sign, this corresponds to a $p$-$n$ junction embedded in neutral graphene.
We define $\ve=\sigma\ns_0 k\ns_0 (d\ns_1+d\ns_2)$, $\theta\ns_y=q (d\ns_1+d\ns_2)$, and $\nu\ns_j=V\ns_j (d\ns_1+d\ns_2)/\hbar\vF$.
We furthermore define $x\ns_j\equiv d\ns_j/(d\ns_1+d\ns_2)$, so $x\ns_1+x\ns_2=1$.
Applying the transfer matrix formalism, we once again arrive at an eigenvalue condition $F(\ve,\nu\ns_1,\nu\ns_2,x\ns_1,\theta\ns_y)=0$,
this time with
\begin{align}
&F(\ve,\nu\ns_1,\nu\ns_2,x\ns_1,\theta\ns_y)=x\ns_1\Big(\theta_y^2-\ve\,(\ve-\nu\ns_1) \Big)   \cdot {\sin\chi\nd_1\over\chi\ns_1}
\cdot \cos\chi\nd_2 \nonumber \\
&\qquad\qquad\qquad + x\ns_2\Big(\theta_y^2-\ve\,(\ve-\nu\ns_2)\Big)\cdot\cos\chi\nd_1\cdot {\sin\chi\nd_2\over\chi\ns_2}\nonumber \\
&\qquad + \Bigg[ x\ns_1 x\ns_2\Big(\theta_y^2-(\ve-\nu\ns_1)(\ve-\nu\ns_2)\Big) \cdot {\sin\chi\nd_1\over\chi\ns_1} \cdot
{\sin\chi\nd_2\over\chi\ns_2}\nonumber\\
& \qquad\qquad\qquad +\cos\chi\nd_1\cos\chi\nd_2\Bigg] \cdot \sqrt{\theta_y^2-\ve^2\>}\ ,
\end{align}
where
\begin{equation}
\chi\nd_j=x\nd_j\sqrt{(\ve-\nu\ns_j)^2-\theta_y^2\>}\ ,
\end{equation}
and where we continue to express energies and wavevectors as dimensionless quantities.  As with the single
trench case, the expression for $F$ is analytically continued to the regimes where $(\ve-\nu\ns_j)^2 < \theta_y^2$.

Generally speaking, when two semi-infinite undoped regions are separated by a sequence of doped slabs, in addition to the Dirac cones
in the bulk one obtains localized subbands with energies lying within the gap between the upper and lower Dirac cones.  The wavefunctions
for these subband states are localized within the slabs.  In Fig. \ref{bands_07}, we plot the Dirac cones and subband dispersions for the case
of a single strip of width $d$ with potential $V=14\,\hbar\vF/d$.  In Figs. \ref{bands_08} and \ref{bands_09}, we investigate the case of
a $p$-$n$ junction embedded in neutral graphene, where both $p$ and $n$ regions have the same width $d$, and again we choose
$V=\pm 14\,\hbar\vF/d$ for purposes of illustration.  Apparently $E=0$ always lies within a gap between subbands.
Note that the parameters of the $p$-$n$ junction are such that the spectum is particle-hole symmetric.

\begin{figure}[t]
\begin{center}
\includegraphics[width=0.45\textwidth]{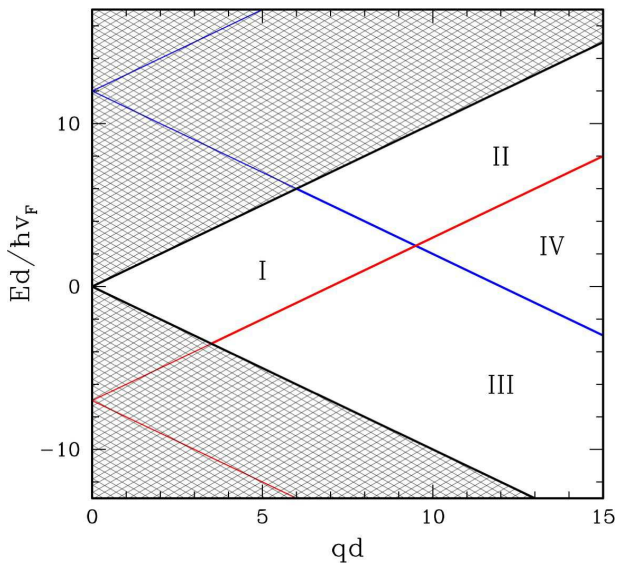}
\end{center}
\caption{The bulk Dirac cones for $V=0$ are hatched in black.   Localized states in the $p$ trench ($V_1=12\hbar v_\ssr{F}/d$) are present in regions
I and III.  Localized states in the $n$ trench ($V_2=-7\hbar v_\ssr{F}/d$) are present in regions I and II.}
\label{dirac_cones}
\end{figure}

\begin{figure}[t]
\begin{center}
\includegraphics[width=0.45\textwidth]{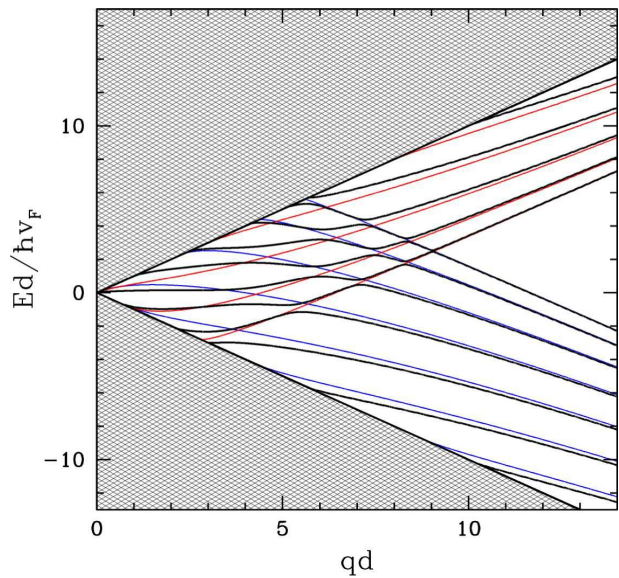}
\end{center}
\caption{Subband structure and bulk Dirac cones for the $p$-$n$ junction with $d_1=d_2=d$ and
$V_1=12\hbar v_\ssr{F}/d$, $V_2=-7\hbar v_\ssr{F}/d$, shown in black.  Superimposed in blue and red are the
subband structures for the individual $V_1$ and $V_2$ barriers, respectively.}
\label{avoided_crossings}
\end{figure}

The subband structure of a general $p$-$n$ junction can be understood in terms of the simple diagram in Fig. \ref{dirac_cones}.
In the figure, we sketch the three Dirac cones for the undoped bulk region (black lines),
the $p$-type slab (blue lines, $V\ns_1 d\ns_1/\hbar\vF=12$), and the $n$-type slab (red lines, $V\ns_2 d\ns_2/\hbar\vF=-7$).
For the purpose of generality, we consider a system which has no particle-hole symmetry.
Outside the black hatching, there are no bulk states.  In regions I and III, there are
localized states living along the $p$-trench.  Similarly, in regions I and II, there are localized states
living along the $n$-trench.  In region I, there can be resonant tunneling between the $p$ and $n$ trenches,
and the localized states live in both trenches.  In region IV, which lies outside all three Dirac cones, there can be no states.
In Fig. \ref{avoided_crossings}, we plot the subband structure for this $p$-$n$ junction, with equal thicknesses
of the $p$ and $n$ regions, superimposing on those curves the results for the individual single barrier problems.
A comparison with the sketch in Fig. \ref{dirac_cones} reveals the basic physics of the energy spectrum.
One may see the confined states associated with each of the two trenches, one set dispersing up
and the other down.  These confined states admix where they are degenerate for the individual
trenches, leading to anticrossings.  In the case where the two trenches are equal and opposite
(e.g., Figs. \ref{bands_08} and \ref{bands_09}), this leads to a set of anticrossings around
zero energy, rather than the extra Dirac points one finds in the superlattice case.

\begin{figure}[t]
\begin{center}
\includegraphics[width=0.45\textwidth]{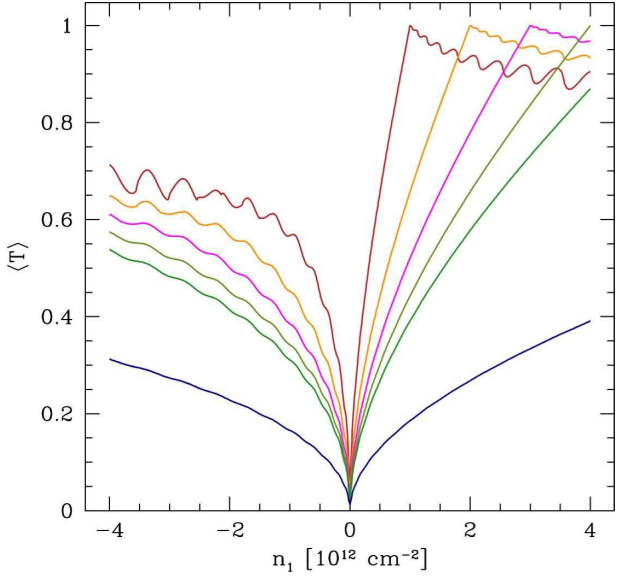}
\end{center}
\caption{Average transmission coefficient across a single trench of width $d_1=100\,$nm as a function of density $n_1$ (in units
of $10^{12}\,{\rm cm}^{-2}$.  From top to bottom: $n_0=1$, $2$, $3$, $4$, $5$, and $20$ ($\times 10^{12}\,{\rm cm}^{-2}$).
Negative densities correspond to a $p$-type (hole-doped) trench.}
\label{trench}
\end{figure}

\section{Two-Probe Landauer Conductance}
\label{conductance}
In this section, we consider transmission through a slab geometry for the more general
case where the graphene leads are doped.
Consider the geometry of Fig. \ref{slabs}.  Define a vertical ($\yhat$-directed) surface $\Sigma$ somewhere
to the right of the $N^{\rm th}$ slab.  The current across that surface will be a sum of three contributions.
First, there will be carriers injected from the left side, with chemical potential $\mu\nd_\ssr{L}$,
moving at angle $\theta\in \big[-\half\pi,\half\pi\big]$, which make it past the slabs with probability $T(E,\theta)$.
These are all right-movers.  Second, there will be carriers from the right lead, with chemical potential $\mu\nd_\ssr{R}$,
moving at an angle $\theta\in \big[\half\pi,\frac{3}{2}\pi\big]$.  Because of this range of $\theta$, all these
are left-movers.  Finally, with probability $R'(E,\theta)$, each of these left-moving carriers reflects off the barrier
region, scattering into a state with $\theta'=\pi-\theta$.  Putting this all together,
\begin{align}
I\nd_\Sigma&=L\nd_\Sigma\,\nhat\nd_\Sigma\!\cdot\!\!\int\limits_{-\infty}^\infty\!\!\!dE\>\cN(E)\!\!\!\!
\int\limits_{-\pi/2}^{\pi/ 2}\!\!\!{d\theta\over 2\pi}\,\bigg\{T(E,\theta)\,\bfJ(E,\theta)\,f(E-\mu\nd_\ssr{L})\nonumber\\
&\quad+ \Big[\bfJ(E,\pi+\theta)+R(E,\pi+\theta)\,\bfJ(E,-\theta)\Big]\,f(E-\mu\nd_\ssr{R})\bigg\}\ ,
\end{align}
where $L\nd_\Sigma$ is the length of the surface and $\cN(E)$ is the density of states,
\begin{equation}
\cN(E)= {g\,|E|\over 2\pi \hbar^2 v^2_\ssr{F}}\ ,
\end{equation}
and $g=4$ accounts for spin and valley degeneracy.  $\bfJ(E,\theta)$ is given by
\begin{equation}
\bfJ(E,\theta)=-e v\nd_\ssr{F}\,{\rm sgn}(E)\,\big(\xhat\,\cos\theta + \yhat\,\sin\theta\big)\ .
\end{equation}
We take $\nhat\nd_\Sigma=\xhat$.  Unitarity of the $S$-matrix yields $T(E,\theta)+R(E,\pi+\theta)=1$.
Expanding in the difference $\mu\nd_\ssr{L}-\mu\nd_\ssr{R}$, we obtain at $T=0$
\begin{equation}
I\nd_\Sigma={e^2\over\pi h}\,V\,\big\langle T(E\nd_\ssr{F})\big\rangle\,
{g\,|E\nd_\ssr{F}|\over\hbar v\nd_\ssr{F}}\,L\nd_y\ ,
\end{equation}
where $V=(\mu\nd_\ssr{L}-\mu\nd_\ssr{R})/e$ is the voltage drop.  The average transmission coefficient
is defined as
\begin{equation}
\big\langle T(E)\big\rangle=\int\limits_{0}^{\pi/ 2}\!\!\!d\theta\ T(E,\theta)\,
\cos\theta\ .
\end{equation}
The carrier density at zero temperature is $n=g k_\ssr{F}^2/4\pi$, so the Fermi energy is
$E\nd_\ssr{F}=\hbar v\nd_\ssr{F} k\nd_\ssr{F}=\hbar v\nd_\ssr{F}\sqrt{4\pi n/g}$.
Thus,
\begin{equation}
I(V)={e^2\over \pi h}\,\big\langle T(E\nd_\ssr{F})\big\rangle\,(4\pi gn)^{1/2}\,L\nd_y\,V\ .
\end{equation}
Defining the effective channel number,
\begin{equation}
N\nd_{\rm c}={g\over\pi}\,k\nd_\ssr{F}L\nd_y=(4gn/\pi)^{1/2}\,L\nd_y\ ,
\end{equation}
the two-terminal Landauer conductance becomes
\begin{equation}
G={I\over V}=N_{\rm c}\times{e^2\over h}\,\big\langle T(E\nd_\ssr{F})\big\rangle\ .
\end{equation}
Expressed in a convenient set of units, the resistance is
\begin{align}
R&=G^{-1}={h\over N_{\rm c} \, e^2}\times
\big\langle T(E\nd_\ssr{F})\big\rangle^{-1}\nonumber\\
&=0.114\,{\rm k}\Omega\cdot\big\langle T(E\nd_\ssr{F})\big\rangle^{-1}
\cdot{\bar n}^{-1/2}\cdot {\bar L}_y^{-1}\ ,
\end{align}
where ${\bar n}$ is the dimensionless carrier density (in the leads) in units of $10^{12}\,{\rm cm}^{-2}$ and
${\bar L}\ns_y$ is the length of the barrier region in units of microns.

\begin{figure}[t]
\begin{center}
\includegraphics[width=0.45\textwidth]{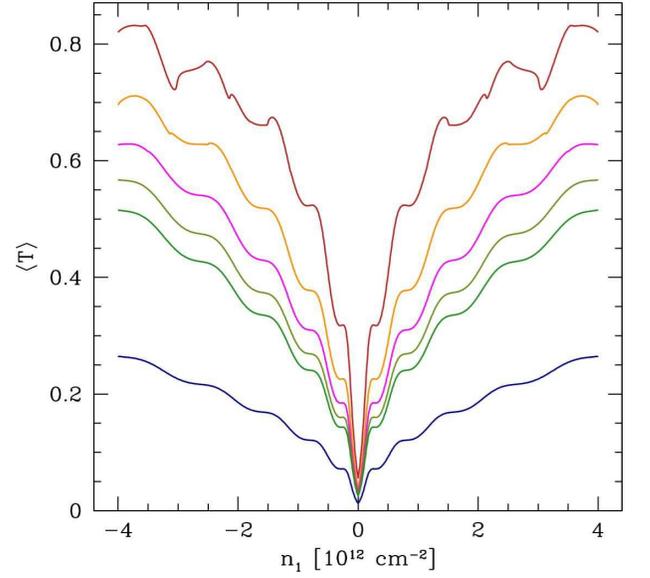}
\end{center}
\caption{Average transmission coefficient across an embedded $p$-$n$ junction with $d_1=d_2=50\,$nm as a function of density $n_1=-n_2$
(in units of $10^{12}\,{\rm cm}^{-2}$.  From top to bottom: $n_0=1$, $2$, $3$, $4$, $5$, and $20$ ($\times 10^{12}\,{\rm cm}^{-2}$).}
\label{embedded}
\end{figure}

For a single trench of width $d\ns_1$ and carrier density $n\ns_1$ embedded between graphene leads of carrier
density $n\ns_0$, we have ${\cal M}=M_0^{-1}\,Q\ns_1\,M\ns_0$, where $Q\ns_j\equiv M\ns_j\,K\ns_j M^{-1}_j$, and
\begin{align}
\textsf{Re}\,{\cal M}\ns_{11}&=\cosh\omega\ns_1\\
\textsf{Im}\,{\cal M}\ns_{11}&=\sigma\ns_0\,\sigma\ns_1 \sinh\omega\ns_1\tan\varphi\ns_1\sec\theta\ns_0\\
&\qquad\qquad\quad - \sinh\omega\ns_1\sec\varphi\ns_1\tan\theta\ns_0 \ .\nonumber
\end{align}
Here we have assumed $k\ns_1 < q$ so the trench contains evanescent solutions, but by a simple analytic continuation
our result also holds for the $k\ns_1 > q$ case.

For an embedded junction consisting of two consecutive slabs $(n\ns_1,d\ns_1)$ and $(n\ns_2,d\ns_2$), the total
transfer matrix is ${\cal M}=M_0^{-1}\,Q\ns_1\,Q\ns_2\,M\ns_0$, resulting in
\begin{align}
&\textsf{Re}\,{\cal M}\ns_{11}=\cosh\omega\ns_1\cosh\omega\ns_2\\
&\quad + \sinh\omega\ns_1\sinh\omega\ns_2\,\big(\sec\varphi\ns_1\sec\varphi\ns_2-\sigma\ns_1\,\sigma\ns_2\,
\tan\varphi\ns_1\tan\varphi\ns_2\big)\nonumber\\
&\textsf{Im}\,{\cal M}\ns_{11}=\sigma\ns_0\,\sigma\ns_1\,\sec\theta\ns_0
\sinh\omega\ns_1\tan\varphi\ns_1\cosh\omega\ns_2\\
&\hskip2.0cm + \sigma\ns_0\,\sigma\ns_2\,\sec\theta\ns_0 \cosh\omega\ns_1
\sinh\omega\ns_2\tan\varphi\ns_2\nonumber\\
&-\tan\theta\ns_0\,\big(\cosh\omega\ns_1\sinh\omega\ns_2\sec\varphi\ns_2
+\sec\varphi\ns_1\sec\varphi\ns_1\cosh\omega\ns_2\big)\nonumber\ .
\end{align}

In Fig. \ref{trench} we plot the average transmission coefficient $\langle T(E\ns_\ssr{F})\rangle$ as a function
of the density $n\ns_1$
for a single trench with $d\ns_1=100\,$nm, for several different values of the lead density $n\ns_0$.   There are three features
worthy of note.  First, there is a nonzero transmission even when $n\ns_1=0$, due to the presence of evanescent solutions
within the trench \cite{TTTRB06}.  Second, there is (obviously) perfect transmission when $n\ns_1=n\ns_0$.  Third, the transmission falls
off rapidly as $n\ns_1\to 0$, and is smooth for $0 < n\ns_1 < n\ns_0$, which is a regime where the carriers in both regions are of the
same sign, yet total internal reflection can occur at the interface, with the critical angle given by $\sin\theta_0^{\rm c}= \big| n\ns_1 / n\ns_0\big|^{1/2}$.
Finally, when $n\ns_1 > n\ns_0$ or when the carriers in the leads and the trench are of opposite sign, one observes quantum oscillations in
the transmission, due to constructive and destructive interference effects from scattering at the walls of the trench.

It is interesting to compare Fig. \ref{trench} with  
results in Ref. \onlinecite{young09}, which reports the measured
conductance through an electrostatically defined trench potential
in a wide graphene ribbon as a function of the gate voltage
defining the trench.  As in our results, the experiments indicate
a rapidly falling conductance when $n\ns_1$ has the
same sign as $n\ns_0$, as $n\ns_1$ approaches zero, and after passing
through zero one observes oscillations due to quantum interference.
A prominent difference is the absence in the experimental
results of a rising background around
which these oscillations occur.  In our model, this appears because the
density of states in the trench increases as $|n\ns_1|$ increases.
It is likely that the difference between our theoretical results and
those observed in Ref. \onlinecite{young09} are due to the relatively
smooth potential induced by the remote gates in the experiment.  In
particular this can yield relatively large ``depletion'' regions between
the trench and the graphene leads which may dominate the resistance of
the device, and overwhelm the increase in conductance related to
the trench density of states.  In principle, one could model the
device as a series of piecewise constant potentials and use the
transfer matrix method we have developed to obtain a more
quantitatively accurate model.

In Fig. \ref{embedded} we plot $\langle T(E\ns_\ssr{F})\rangle$ as a function of
$n\ns_1$ for  an embedded $p$-$n$ junction
with $d\ns_1=d\ns_2=50\,$nm and $n\ns_2=-n\ns_1$.  For such a symmetric junction the average transmission is symmetric under
$n\ns_1 \to -n\ns_1$.  The step-like features present in the transmission here represent
scattering effects due to the localized solutions and the avoided crossings in their
dispersions, as discussed in the previous section.  Thus the bound
states support an (in principle) measurable signature in conductance through the system.


\section{Summary}
\label{summary}
In this paper we have studied the spectra of and transport properties through
graphene in the presence of several piecewise constant potentials.  We demonstrated
that these may be investigated in a straightforward manner using a transfer
matrix approach.  For the case of superlattice potentials, we demonstrated
that exact solutions of the Dirac equation generically contain Dirac points
centered at zero energy for particle-hole symmetric potentials \cite{brey09,park09},
and that these are special cases of Dirac points that appear
throughout in the band structure of superlattice potentials.
We also considered the cases of a single trench and an embedded
$p$-$n$ junction in neutral graphene, both of which may support confined
states at large enough values of $k_y$, the momentum along the
translationally invariant direction.  An analysis of conductance
through these structures shows quantum interference effects which
may be interpreted as signatures of these bound states, in principle
allowing these unique features of graphene nanostructures to be seen
experimentally.

{\sl Note added:} While this manuscript was in the final stages of preparation, a manuscript by Barbier, Vasilopoulos, and Peeters
appeared \cite{barbier10} which also considers the effect of a superlattice on monolayer graphene and the emergence of additional Dirac points.
We are grateful to F. M. Peeters for subsequently alerting us to the substantial previous work which he and his collaborators have contributed
on the subject of Dirac subband spectra and transport in the presence of one-dimensional potentials\cite{pereira07,barbier08,barbier09}.  We are also
grateful to Y. P. Bliokh for drawing our attention to Ref. \onlinecite{bliokh09}, which also addresses subband formation in the presence of a
superlattice structure.

\section{Appendix}
We define $Q=M K M^{-1}$ for a given slab.
Within a propagating slab, we have
\begin{equation}
Q=\begin{pmatrix} \cos\chi + \sin\chi \tan\theta & i\sigma\sin\chi\sec\theta \\
i\sigma\sin\chi\sec\theta & \cos\chi - \sin\chi \tan\theta \end{pmatrix}\ ,
\end{equation}
with $\chi=kd\cos\theta$, while for a nonpropagating slab, we have
\begin{equation}
Q=\begin{pmatrix} \cosh\omega  + \sinh\omega\sec\varphi& i\sigma\sinh\omega\tan\varphi \\
i\sigma\sinh\omega\tan\varphi & \cosh\omega - \sinh\omega\sec\varphi \end{pmatrix}
\end{equation}
with $\omega=kd\cot\varphi$.   These expressions are useful when constructing the transfer matrix for a series of slabs,
which is given in Eq. \ref{transfermat}.

\section{Acknowledgements}
DPA thanks M. Fogler for helpful discussions.  Funding was provided
by the MEC-Spain via Grant No. FIS2009-08744 (LB), and by the NSF through
Grant No. DMR-0704033 (HAF) and EEC-0646547 (EAK).   DPA, LB, HAF, and KZ acknowledge the
hospitality of the Benasque Physics Center (Spain), where this work was initiated.

\vfill\eject


\bibliography{ONED.bib}


\end{document}